\documentclass[aps,prd,amsfonts,amssymb,preprint,
eqsecnum,nofootinbib,superscriptaddress]{revtex4}
\usepackage{amsmath,bbm,mathrsfs}
\usepackage{hyperref,graphicx,subfigure}

\newcommand\tm{\tilde{m}}
\begin{document}

\title{
Nonlinear Boundary Dynamics and Chiral Symmetry in Holographic QCD
\vskip 0.1in}
\author{Dylan Albrecht}
\author{Joshua Erlich}
\affiliation{High Energy Theory Group, Department of Physics,
College of William and Mary, Williamsburg, VA 23187-8795}
\author{Ronald J. Wilcox}
\affiliation{Department of Mechanical Engineering, Massachusetts Institute of Technology,
Cambridge, MA 02139}

%

\newcommand\sect[1]{\emph{#1}---}

\begin{abstract}
In a hard-wall model of holographic QCD, we find that
nonlinear boundary dynamics are required in order 
to maintain the correct pattern of
explicit and spontaneous chiral symmetry breaking
beyond leading order in the pion fields.
With the help of a field redefinition, 
we relate the requisite nonlinear boundary conditions to a standard
Sturm-Liouville system.
Observables insensitive to the chiral limit 
receive only small corrections in the improved description, and classical 
calculations in the hard-wall model remain surprisingly accurate.

\end{abstract}
\maketitle

\section{Introduction}
Holographic QCD is an extra-dimensional approach to modeling hadronic physics
at low energies 
\cite{Csaki:1998qr,Polchinski:2000uf,Brodsky:2003px,deTeramond:2005su,Erlich:2005qh,Da Rold:2005zs}.  
Hadronic resonances are interpreted as Kaluza-Klein
modes of five-dimensional (5D) fields with quantum numbers of the corresponding
QCD states. Motivated by the AdS/CFT correspondence
\cite{Maldacena:1997re,Witten:1998qj,Gubser:1998bc}, 
in hard-wall models 
\cite{Polchinski:2000uf,Brodsky:2003px,deTeramond:2005su,Erlich:2005qh,Da Rold:2005zs} 
the background is chosen
to be a slice of Anti-de Sitter space (AdS$_5$), with metric
\begin{equation}
ds^2=\frac{R^2}{z^2}\left(\eta_{\mu\nu}dx^\mu dx^\nu-dz^2\right),\ \epsilon<z<z_m,
\label{eq:AdSmetric}
\end{equation}
where $\eta_{\mu\nu}$ is the Minkowski metric with mostly negative signature
in 3+1 dimensions, \(R\) is the AdS radius, and $\epsilon$ provides a
short-distance cutoff in the 
model.  The size of the extra dimension depends on $z_m$, which sets the 
Kaluza-Klein scale and is related holographically to the confining scale of QCD.
Alternative spacetime backgrounds have been motivated by
D-brane configurations in string theory which give rise to QCD-like theories
with chiral symmetry breaking and confinement, as in the Sakai-Sugimoto model
based on the D4-D8 system \cite{Sakai:2004cn}.
Non-normalizable ({\em i.e.}~infinite-action) 
backgrounds of fields act as sources of corresponding
operators in QCD, and normalizable ({\em i.e.}~finite-action) 
backgrounds determine the expectation values of those operators
\cite{Balasubramanian:1998sn,Klebanov:1999tb}.
The established AdS/CFT dictionary between physics in 3+1 and 4+1 dimensions 
motivates a model in which a complex scalar field in 4+1 dimensions, 
with the quantum numbers of the
quark bilinears $q_{Li}\overline{q}_{Rj}$ with flavor labels $i,j$, 
fluctuates about a background 
configuration related to the quark mass  (the source of $\overline{q}q$)
and chiral condensate (the expectation value of $\overline{q}q$).  We will
review this version of the hard-wall model in more detail in 
Sec.~\ref{sec:hardwall}.
As a demonstration of the pattern of chiral symmetry breaking in the model, 
the Gell-Mann-Oakes-Renner (GOR) relation
for the pion mass was shown to be approximately satisfied
\cite{Erlich:2005qh,Erlich:2008gp}, 
\begin{equation}
m_\pi^2 f_\pi^2 = 2 m_q \sigma, \label{eq:GOR} \end{equation}
where $m_\pi$ and $f_\pi$ are the pion mass and decay constant calculated
from the model, and $m_q$ and $\sigma$ are  parameters in the scalar field
background 
playing the role of the quark mass and chiral condensate, respectively.  
The GOR relation is also satisfied in an SU(3) extension of the hard-wall
model with an independent strange quark mass parameter \cite{Abidin:2009aj}.

Classical calculations in the hard-wall model have reproduced
a variety of QCD observables with surprising accuracy, generally at the 
10-15\% level \cite{Erlich:2005qh,Da Rold:2005zs}.  The hard-wall model fails at high energies
\cite{Csaki:2008dt}
where the Kaluza-Klein spectrum diverges from the Regge spectrum 
\cite{Shifman:2005zn}, a problem partially corrected in the soft-wall model
\cite{Karch:2006pv}.  
More surprisingly,
it was discovered that pion condensation in the hard-wall model  
has qualitatively different features from predictions of the chiral
Lagrangian \cite{Albrecht:2010eg}.  If the isospin chemical potential increases 
beyond a critical value,
hadronic matter is expected to undergo a phase transition to
a state in which a linear combination of the  pion fields condenses
\cite{Migdal:1976zc,Sawyer:1973fv,Kaplan:1986di}.
In chiral perturbation theory without unphysically 
large low-energy coefficients, the pion condensation transition 
is second order and approaches the Zel'dovich equation
of state for stiff matter
smoothly across the phase boundary \cite{Son:2000xc}.  The pion condensation
transition has been studied using other approaches including lattice QCD 
\cite{Kogut:2002zg,deForcrand:2007uz,Detmold:2008fn,Detmold:2008yn}, 
with results in agreement with chiral perturbation theory.
However, in the hard-wall model the transition from the hadronic
phase to the pion condensate
phase was found to be first order, rapidly approaching the 
Zel'dovich equation of state across the phase boundary.  The holographic
model takes as input the pattern of chiral symmetry breaking, so
disagreement with lowest-order chiral perturbation theory is surprising.

Another more subtle puzzle related to chiral symmetry lies in the form of the GOR
relation when the chiral condensate $\sigma$ is allowed to have a phase with respect
to the quark mass.  As deduced from a linear sigma model, the quark mass and 
condensate are in phase, 
a fact related to the existence of an anomalous chiral U(1) symmetry.  However, 
beginning with the non-Abelian nonlinear sigma model we can ask
what would happen if the parameter $\sigma$ corresponding to
$\langle \overline{q}_Lq_R\rangle$ were complex.  In unpublished work, it was found that
the resulting GOR relation as derived in the hard-wall model with complex $\sigma$
disagrees with the analogous prediction of the chiral Lagrangian.
In particular, in the hard-wall model the GOR relation takes the form 
\cite{Ron-thesis}
\begin{equation}
m_\pi^2 f_\pi^2 \log\left(\frac{\sigma}{\sigma^*}\right)=2 m_q\left(\sigma-
\sigma^*\right),\end{equation}
while the corresponding
prediction based on the chiral Lagrangian and PCAC is the same as Eq.~(\ref{eq:GOR})
with $\sigma$ replaced by its real part.
It is the goal of this paper to reconcile these discrepancies and restore 
consistency between holographic QCD and chiral perturbation theory.

We will show that the incorrect structure of pion interactions in
the hard-wall model is a result of the choice of 
boundary conditions imposed on the 5D fields.
Care must be taken in order that the
boundary conditions respect the symmetry breaking structure.
The subtlety compared with other
extra-dimensional models is that here the background of a bulk field not only
spontaneously breaks the gauge invariance in the bulk, but a non-normalizable
term in the background also explicitly breaks the gauge invariance.  In the hard-wall model
the bulk gauge invariance is responsible for the global chiral symmetry of the
effective 4D theory, so it is important that the explicit and spontaneous
breaking be correctly accounted for at the boundary.  

We will demonstrate that nonlinear boundary conditions on the bulk scalar 
field, or else nonlinear boundary terms in the action, may be 
consistently chosen so as to
restore the proper pattern of chiral symmetry breaking in the hard-wall model.
The nonlinear boundary dynamics we propose are 
an alternative to the description in
Ref.~\cite{Domenech:2010aq}, which also leads to acceptable symmetry structure
(and also accommodates a bulk Chern-Simons term absent in the present model).
The nonlinear boundary conditions relate the 5D $\sigma$-model scalars
to products of pseudoscalars $\pi^a \pi^a$, $(\pi^a \pi^a)^2$, {\em etc}.
In order to motivate these unusual boundary conditions and to demonstrate
their relation to a Sturm-Liouville system, as 
required for consistency of the standard
Kaluza-Klein decomposition of the fields and their interactions, 
we reparametrize the 5D
fields by a nonlinear field redefinition.  
The reparametrization introduces a new surface term ({\em i.e.}~a
total derivative) involving the pions
in the 5D action but replaces the nonlinear boundary conditions 
with ordinary linear boundary conditions
consistent with the desired symmetry-breaking pattern.  As opposed to the
boundary conditions proposed to
describe multitrace operators in the AdS/CFT correspondence 
\cite{Witten:2001ua,Berkooz:2002ug}, 
the nonlinear boundary conditions in the hard-wall model arise away from
the boundary of AdS, at the infrared boundary of the 
spacetime.

The modifications of the hard-wall model as described in 
Ref.~\cite{Erlich:2005qh} required to restore the structure
of chiral symmetry breaking have a number of phenomenological consequences.
The GOR relation for the pion mass, Eq.~(\ref{eq:GOR}), is correctly
normalized only after the quark mass and chiral condensate are rescaled.  
This same rescaling is consistent with the AdS/CFT correspondence, and is
a result of the modified boundary physics.
The pion
potential is modified so as to reconcile properties of the pion condensation
transition with predictions of chiral perturbation theory.  Most hadronic 
observables receive only small
corrections which vanish in the chiral limit, so the hard-wall model
remains surprisingly accurate in many of its predictions for low-energy
QCD observables.

\section{Review of the Hard-Wall Model}
\label{sec:hardwall}
Following the conventions of Ref.~\cite{Erlich:2005qh}, 
the hard-wall model is defined by the 5D action \begin{equation}
{\cal S}=
\int d^5x \sqrt{|g|} {\rm Tr}\left\{-\frac{1}{4g_5^2}\left((F_{MN}^{(L)})^2+
(F_{MN}^{(R)})^2\right)+|D_MX|^2+\frac{3}{R^2}|X|^2\right\},
\label{eq:5Daction}\end{equation}
where $F^{(L)}=F^{(L)a}\sigma^a/2$ and $F^{(R)}=F^{(R)a}\sigma^a/2$ are field strengths
for the 5D SU$(2)_L\times$SU$(2)_R$ gauge fields; $\sigma^a$ are
the three Pauli matrices; $M,N\in 0,\dots,4$ are Lorentz indices contracted
with the AdS$_5$ metric from Eq.~(\ref{eq:AdSmetric}); and
$X$ is a 2$\times$2 matrix of complex scalar fields transforming
in the bifundamental representation of SU$(2)_L\times$SU$(2)_R$.  For the
calculations in this paper we work with \(R = 1\).

The equations of motion have a solution with vanishing gauge fields and
scalar field profile
\begin{equation}
X_0(z)=\frac{1}{2}(\tm_q z+\sigma z^3)\openone\equiv \frac{\tilde{v}(z)}{2}\openone, 
\label{eq:X0}\end{equation}
where $\openone$ is the 2$\times$2 identity matrix.
The fields $X$ have the quantum numbers of the scalar quark bilinears, which
are the operators whose coefficients in the Lagrangian of
the 3+1 dimensional theory are quark masses.  
We approximate isospin as unbroken, so that
the up and down quarks have equal mass.
The term in the solution Eq.~(\ref{eq:X0}) 
proportional to $z$ is non-normalizable and is related by
the AdS/CFT dictionary to
the quark mass, which explicitly breaks the chiral symmetry; and the term in
the solution
proportional to $z^3$ is normalizable and is related to the condensate
$\langle q_L\overline{q}_R\rangle$, which spontaneously breaks the chiral 
symmetry \cite{Balasubramanian:1998sn,Klebanov:1999tb}.

The non-normalizable mode in the scalar field background explicitly breaks
a bulk gauge invariance, but the presence of this mode is equivalent to 
spontaneous breaking due to a heavy Higgs field
localized on the ultraviolet boundary ($z=\epsilon$) in the decoupling
limit.  To see this, we write the Higgs doublet $(\phi^+,\phi^0)$ as a
matrix, \begin{equation}
H=\left(\begin{array}{cc}
\overline{\phi^0}& \phi^+ \\
-\phi^-&\phi^0 \end{array}\right), \end{equation}
which transforms in the bifundamental representation of the chiral symmetry.
In this form, the up and down quark Yukawa couplings take the form
${\cal L}_{\rm Yuk}=
-\lambda {\rm Tr}\left\{H^\dagger q_L \overline{q}_R+h.c.\right\}$.  
Replacing $(-)q_L \overline{q}_R$ with the 5D field $X$, the localized Higgs
field has boundary action
\begin{equation}
{\cal S}_\epsilon= \int_{z=\epsilon} d^{4}x\, {\rm Tr}\left\{ 
\left| \partial_\mu H \right|^2  - V(H)    + 
\frac{\lambda}{\epsilon^{3}} \left( H X^{\dagger}
      + X H^{\dagger}\right)\right\},
\label{eq:Higgsboundary}\end{equation}
where \(V(H)\) is the Higgs potential exhibiting spontaneous
symmetry breaking. A similar coupling appears in bosonic technicolor models
\cite{Carone:1993xc,Carone:2006wj}.
The factor of \(1/\epsilon^3\) in the last term ensures
the proper scaling with the field $X$ near the UV boundary.  
We impose Neumann boundary conditions on $X$ in the ultraviolet, modified by
the presence of the boundary term (\ref{eq:Higgsboundary}).
Replacing the Higgs field $H$ by its vacuum expectation value $\langle H
\rangle=\langle \phi^0 \rangle \openone$, chosen real, 
the equations of motion and boundary condition for $X$ are given by:
\begin{equation}
\partial_{z}\left( \frac{1}{z^{3}} \partial_{z} X \right) - \frac{1}{z^{3}}
  \Box X + \frac{3}{z^{5}} X = 0, 
\label{eq:XHeom} \end{equation}
\begin{equation}
  \partial_{z} X \big|_{\epsilon} = -\lambda \langle H \rangle,
\end{equation}
where \(\Box \equiv \eta^{\mu \nu} \partial_{\mu} \partial_{\nu}\).
Greek indices will always refer to 3+1 dimensions, and capital Latin
indices will refer to 4+1 dimensions.

By identifying the diagonal quark mass
\(M_{q} = -\lambda \langle H \rangle\), the boundary condition becomes
\(\partial_{z} X \big|_{\epsilon} = M_{q}\).  Near the boundary as $\epsilon
\rightarrow 0$,
the solution for $X$ consistent with this boundary condition 
takes the form $X\approx M_q z$.
Thus, the coupling of $X$ 
to a Higgs field localized at the UV boundary gives rise to the appropriate 
non-normalizable background solution for $X$, which justifies the 
presence of the non-normalizable background and its AdS/CFT
interpretation as the source for the operator $q\overline{q}$.
However, the overall normalization of $M_q$ in terms of the physical quark mass
depends on the normalization
of the field $X$, and may also be modified by infrared dynamics as will be
the case here.  
In Eq.~(\ref{eq:X0}) we have set $M_q=\tm_q/2$, although other normalizations 
better match QCD predictions for correlators of products of scalar
quark bilinears \cite{Cherman:2008eh}.

The fluctuations of $X$, which contain scalars and pseudoscalars (pions),
are typically decomposed as \cite{Erlich:2005qh,Da Rold:2005zs,DaRold:2005vr}: \begin{equation}
X_{old}(x,z)=\left(\frac{1}{2}
(\tm_q z +\sigma z^3)\openone + \tilde{S}(x,z)\right)\tilde{U}(x,z),
\label{eq:Xold} \end{equation}
where $\tilde{S}$ is a Hermitian matrix of scalars
and $\tilde{U}= {\rm exp} [i \tilde{\pi}^{a}(x, z) \sigma^{a}]$ is unitary.  Any
matrix can be written as a product of a Hermitian and a unitary matrix, and
any Hermitian matrix function of $x$ and $z$ can be written as the term in
parentheses in Eq.~(\ref{eq:Xold}), so this ansatz is completely general up
to a U(1) factor relevant for the chiral anomaly but which will not be 
discussed here.

The scalars and pseudoscalars decouple at quadratic order in the action, so
we temporarily limit our attention to fluctuations with $\tilde{S}=0$.  
In order to simplify the discussion we also temporarily decouple the
vector fields by taking $g_5=0$.  We will include the gauge couplings
in Sec.~\ref{sec:g5}, but they are an added complication which is not necessary to
understand the main conclusions.

The lightest pion Kaluza-Klein mode, $\tilde{\pi}(x,z)=
\tilde{\pi}^a(x,z)\sigma^a/2=\tilde{\pi}^a(x)\psi(z)\sigma^a/2$, has
action
\begin{equation}
\begin{split}
{\cal S} &= \int d^{5}x \,{\rm Tr}\left\{\frac{\tilde{v}(z)^2}{4z^3}
    (\partial_{\mu} \tilde{U} \partial^{\mu}\tilde{U}^{\dagger} - \partial_{z}
    \tilde{U} \partial_{z} \tilde{U}^{\dagger})
      \right\}\\
&= \int d^{5}x \,{\rm Tr} \left\{ \frac{\tilde{v}(z)^2}{4z^3} \left(\partial_{\mu}
    \tilde{U} \partial^{\mu}\tilde{U}^{\dagger} -  4 \psi'(z)^2\,\tilde{\pi}(x)^2
    \right)
      \right\}.
\label{eq:Sold}\end{split}\end{equation}
As explained in Ref.~\cite{Erlich:2005qh} and will also be explained in 
Sec.~\ref{sec:chisym}, 
the pion wavefunction is flat with $\psi(z)\approx 1$ 
except near $z=\epsilon$, so integrating over $z$ yields the effective 4D 
action for the pions, \begin{equation}
{\cal S}_{eff} = \int d^4x \, \frac{f_\pi^2}{4}{\rm Tr}\left\{\partial_{\mu}
    \tilde{U} \partial^{\mu}\tilde{U}^{\dagger} - 4 m_\pi^2 \tilde{\pi}^2\right\},
\label{eq:Seffold}\end{equation}
where $m_\pi^2$ is determined by the equations of motion and boundary conditions, and
from the kinetic term we identify the pion decay constant
\begin{equation}
\begin{split}
f_\pi^2&\approx\int_\epsilon^{z_m}dz\, \sigma^2 z^3 \\
&= \frac{\sigma^2 z_m^4}{4}
\label{eq:fpi}\end{split}\end{equation}
as $\epsilon\rightarrow0$.
The expression (\ref{eq:fpi})
for $f_\pi$ also follows from an AdS/CFT calculation of the transverse
part of the axial vector current-current correlator \cite{Albrecht:2010eg}.

\section{Chiral Symmetry Breaking in the Hard-Wall Model}
\label{sec:chisym}
The structure of the pion effective action (\ref{eq:Seffold}) 
demonstrates the discrepancy between the hard-wall model as defined above
and chiral perturbation theory.
The pion mass term in Eq.~(\ref{eq:Seffold}) does not include the higher-order pion 
interactions required for the chiral symmetry to be maintained while the quark mass 
matrix transforms in the bifundamental representation  
under the chiral symmetry (like a Higgs spurion).
In the chiral Lagrangian the pion mass term is proportional to ${\rm Tr}\,(M_q U^\dagger
+M_q^\dagger U)$, which displays the proper pattern of explicit and spontaneous 
chiral symmetry breaking.  Beyond quadratic order in the pion fields,
the hard-wall model as described above
disagrees with the chiral Lagrangian, 
leading to unusual pion phenomenology inconsistent
with chiral perturbation theory. The absence of quartic terms in the pion
potential in this context was also noted in Ref.~\cite{Kelley:2010mu}.

Restoration of the correct pattern of chiral symmetry breaking may be achieved by
modifying the boundary conditions in a nonlinear way which mixes 
the scalar modes $\tilde{S}$ and products of pseudoscalars $\tilde{\pi}^a\tilde{\pi}^a$, 
as we discuss below.  It will be convenient to 
rescale the quark mass parameter $\tm_q=-2m_q$, so that the background profile of the
field $X$ takes the form \begin{equation}
X_0(z)=-m_qz +\frac{\sigma z^3}{2}\equiv \frac{v(z)}{2}\openone. 
\end{equation}
We then consider a nonlinear redefinition of the
5D fields as follows:
\begin{equation}
X(x,z)=-m_q z +\left(\frac{\sigma}{2}z^3+S(x,z)\right)U(x,z),
\label{eq:Xnew}
\end{equation}
which is to be compared with Eq.~(\ref{eq:Xold}).  We write $U(x,z)=
\exp(i\pi^a(x,z)\sigma^a)$.  Now the pseudoscalar fluctuations in 
$U(x,z)$ multiply the term in the background responsible for the spontaneous breaking
of the chiral symmetry, but not the term responsible for the explicit breaking.  With
boundary conditions $S(x,\epsilon)=\pi^a(x,\epsilon)=S(x,z_m)=0$, and
a Neumann condition on $\pi^a$ at $z_m$, 
the scalar and pseudoscalar modes again decouple
and the pion action takes the form
\begin{multline}
{\cal S} = \int d^{5}x \, {\rm Tr}\left\{\frac{\sigma^2 z^3}{4} (\partial_{\mu}
    U \partial^{\mu}U^{\dagger} - \partial_{z} U \partial_{z} U^{\dagger})
      \right\}
    + \int d^{4}x \, {\rm Tr} \left\{\frac{m_{q} \sigma}{2} (U
    + U^{\dagger})\big|_{z_{m}}\right\},
\label{eq:Swbt}\end{multline}
where the last term is an IR localized boundary term due to a total derivative in 
the action.  
The field parametrization Eq.~(\ref{eq:Xnew}) is an alternative to those of
Refs.~\cite{Panico:2007qd,Domenech:2010aq} which also lead to an acceptable 
model, but with a nonlinear
term at the UV boundary $z=\epsilon$ rather than at the IR boundary $z=z_m$. 
The Kaluza-Klein modes are solutions to the linearized equations of motion,
which follow from the quadratic part of the action:
\begin{equation}
{\cal S} = \int d^{5}x \, \frac{\sigma^2 z^3}{2} (\partial_{\mu}
    \pi^{a} \partial^{\mu} \pi^{a} - \partial_{z} \pi^{a} \partial_{z} \pi^{a})
    - \int d^{4}x \, m_{q} \sigma (\pi^{a} \pi^{a})\big|_{z_{m}},
\end{equation}
where \(\pi^{a}(x,z) = \pi^{a}(x) \psi(z)\).
The linearized equation of motion for the pion wavefunction is
\begin{equation}
\partial_{z} (z^{3} \partial_{z} \psi) = -m_{\pi}^{2} z^{3} \psi.
\end{equation}
The Neumann boundary condition in the IR is modified by the boundary term in the
action, with the result
\begin{equation}
\partial_{z} \psi(z) \big|_{z_{m}} = -\frac{2 m_{q}}{\sigma z_{m}^{3}}
  \psi(z_{m}).
\label{eq:psibdy}
\end{equation}

Note that the boundary conditions here are linear, and the nonlinear boundary
terms in the action are treated as interactions.  In this form, the pions are
described by a standard Sturm-Liouville system. The solutions are in terms of
Bessel functions and the normalizable solution has the expansion
\begin{equation}
\psi(z) = \left(1 - \frac{m_{\pi}^{2} z^{2}}{8}
    + \textrm{higher order in } m_{\pi}z \right).
\end{equation}
If $m_\pi z_m\ll1$ then $\psi(z)\approx 1$ in the entire interval $\epsilon<z<z_m$.
Substituting the expansion of \(\psi(z)\) into the boundary condition 
Eq.~(\ref{eq:psibdy}), we find
to leading order in $m_q$,
\begin{equation}
m_{\pi}^{2} \frac{\sigma^{2} z_{m}^{4}}{4} = 2 m_{q} \sigma.
\label{eq:mpi}\end{equation}
Using Eq.~(\ref{eq:fpi}), which is not affected by the field redefinition,
Eq.~(\ref{eq:mpi}) is just the Gell-Mann-Oakes-Renner relation
\begin{equation}
m_\pi^2 f_\pi^2= 2 m_q \sigma, \end{equation}
justifying the interpretation of $m_q$ and $\sigma$ as the quark mass and chiral 
condensate, respectively, up to a simultaneous rescaling of $m_q$ and $\sigma$ as
in Ref.~\cite{Cherman:2008eh}.  Note that the quark mass, and in particular the product
$m_q\sigma$, is rescaled from the old parameter
$\tm_q$ and even has a different sign.  This rescaling is required in order to obtain 
the correct normalization in the GOR relation, but is also consistent with the AdS/CFT
interpretation of $m_q$ as the source for the operator $(-)q_L\overline{q}_R$ whose
expectation value is the chiral condensate.  The
condensate is obtained by varying the action with respect to
the source $m_q$.  Because of the additional boundary term, which scales as $m_q$, the
quark mass needs to be rescaled with respect to the chiral condensate as above. 

We now derive the 4D effective Lagrangian for the redefined pions.  Let us first
focus on the \(z\)-derivative piece.  Ignoring the higher KK modes and writing
\(U = {\rm exp} (i \pi^{a}(x)\psi(z)\,\sigma^{a})\), we find
\begin{equation}
\int d^{5}x \, {\rm Tr} \left(\frac{\sigma^{2} z^{3}}{4}
    \partial_{z} U \partial_{z} U^{\dagger} \right)
  = \int d^{4}x \, \left[\int dz \,\frac{\sigma^{2} z^{3}}{2}
    \psi'(z)^2 \right] \pi^{a}(x) \pi^{a}(x),
\end{equation}
as before. Integrating by parts and using the equations of motion and the boundary condition
for \(\psi\) we arrive at:
\begin{equation}
{\cal S} \supset \int d^{4}x \left[\int dz \, (-m_{\pi}^{2}
    \sigma^{2} z^{3}) \psi(z)^{2}
  + 2 m_{q} \sigma \right] \frac{1}{2}\pi^{a}(x) \pi^{a}(x).
\end{equation}
Using the flatness of the \(\psi(z)\) profile, the expression for
\(f_{\pi}^{2}\) in Eq.~(\ref{eq:fpi}), 
and the GOR relation, the above expression in brackets vanishes.
Including the boundary term in Eq.~(\ref{eq:Swbt}), the approximate 4D effective 
Lagrangian is equivalent to the lowest-order chiral Lagrangian:
\begin{equation}
{\cal S} = \int d^{4}x \, {\rm Tr} \left\{\frac{f_{\pi}^{2}}{4}\partial_{\mu}U
    \partial^{\mu}U^{\dagger} + \frac{m_{\pi}^{2} f_{\pi}^{2}}{4}
    (U + U^{\dagger})\right\}.
\end{equation}
As a consequence, the properties of
the pion condensate phase and other aspects of pion physics now agree with the
predictions of chiral perturbation theory thanks to the modified boundary
dynamics.

\section{Nonlinear Boundary Conditions}
The nonlinear 
reparametrization of the bulk field $X$ in Eq.~(\ref{eq:Xnew})
allows for independent linear boundary conditions on the scalar and pseudoscalar
modes while maintaining the proper pattern of chiral symmetry breaking.  In terms of the 
original decomposition of $X$ as per Eq.~(\ref{eq:Xold}), the boundary conditions 
required to maintain the pattern of chiral symmetry  mix the
scalar and pseudoscalar fields in a nonlinear way.  To understand the structure of
the nonlinear boundary conditions we can expand the two field decompositions, Eqs.
(\ref{eq:Xold}) and (\ref{eq:Xnew}):
\begin{equation}\begin{split}
X&=\left[\frac{1}{2}(\tm_q z+\sigma z^3)+\tilde{S}\right]\left(1+2i\tilde{\pi}-
2\tilde{\pi}^2+\dots\right) \\
&=-m_q z+\left(\frac{\sigma z^3}{2}+S\right)\left(1+2i\pi-2\pi^2+\dots\right),
\label{eq:fieldredef}
\end{split}\end{equation}
where $\tilde{\pi}=\tilde{\pi}^a\sigma^a/2$, and similarly for $\pi$.
Equating the anti-Hermitian parts of the two descriptions gives, 
to quadratic order in the fields, 
\begin{equation}
\begin{split}
\pi&=\tilde{\pi}\left(1+\frac{\tm_q}{\sigma z^2}\right)+
\frac{1}{\sigma z^3}\left[(\tilde{S}\tilde{\pi}-S\pi)+
(\tilde{\pi}\tilde{S}-\pi S)\right]+\dots \\
&=\tilde{\pi}\left(1+\frac{\tm_q}{\sigma z^2}\right)+
\frac{1}{\sigma z^3}\left\{\left[\tilde{S}\tilde{\pi}-\tilde{S}\tilde{\pi}\left(
1+\frac{\tm_q}{\sigma z^2}\right)\right]+\textrm{h.c.}\right\}+\dots \\
&=\tilde{\pi}\left(1+\frac{\tm_q}{\sigma z^2}\right)-
\frac{\tilde{m_q}}{\sigma^2 z^5}\left(\tilde{S}\tilde{\pi}+\tilde{\pi}\tilde{S}\right)
+\dots.
\end{split}
\end{equation}
Similarly, the Hermitian parts give,
\begin{equation}
\begin{split}
S&=\tilde{S}-\tilde{m}_q z\,\tilde{\pi}^2 -\sigma z^3\left(\tilde{\pi}^2
-\pi^2\right)+i\left[\tilde{S}\left(\tilde{\pi}-\pi\right)-\textrm{h.c.}\right]+\dots
\\
&=\tilde{S}+\tilde{m}_q z\,\tilde{\pi}^2-i\frac{\tm_q}{\sigma z^2}\left(
\tilde{S}\tilde{\pi}-\tilde{\pi}\tilde{S}\right)+\dots.  
\end{split}\end{equation}
These expressions have been left in terms of the old mass parameter $\tm_q$, 
which is equivalent to $-2m_q$, 
and terms higher order in $\tm_q$ have been dropped.

In the new decomposition of the field $X$, the boundary conditions  consistent
with the chiral symmetry breaking structure are: \begin{equation}
S(x,\epsilon)=S(x,z_m)=0=\pi(x,\epsilon)=0,\ \ \ \partial_z\pi(x,z)|_{z_m}=
\frac{\tm_q}{\sigma z_m^3}\pi. \end{equation}
In terms of the original decomposition of the field $X$, the boundary conditions are,
\begin{equation}
S=0\  \rightarrow\ \tilde{S}=-\tm_q z\,\tilde{\pi}^2 +\dots,
\label{eq:nonlinbc}
\end{equation}
\begin{equation}
\partial_z\pi=\frac{\tm_q}{\sigma z^3}\pi
\ \rightarrow\ \partial_z\tilde{\pi}=\frac{3\tm_q}{\sigma z^3}\tilde{\pi}
+\dots,
\end{equation}
where the ellipses include terms higher order in the fields and in $\tm_q$, 
and terms that vanish when traced over.

For these boundary conditions to be physically acceptable,
the boundary variation of the action must vanish.  
Expanding the action (\ref{eq:5Daction}) with $g_5=0$ about the background
as in Eq.~(\ref{eq:Xold}), we obtain
\begin{equation}
\begin{split}
{\cal S} = \int d^5x\,{\rm Tr}\Bigg\{&\frac{1}{z^3}\partial_\mu X
    \partial^\mu X^\dagger
  - \frac{1}{z^3}(\partial_z\tilde{S})^2
  - \frac{\tilde{v}'(z)}{z^3}\partial_z\tilde{S}
  - \frac{1}{z^3}\left(\frac{\tilde{v}}{2}+\tilde{S}\right)^2
    \partial_z \tilde{U}\partial_z \tilde{U}^\dagger \\
   &
  + \frac{3}{z^5}\left(\tilde{v}\tilde{S}+\tilde{S}^2\right)\Bigg\}
  + \textrm{constant}.
\end{split}
\label{eq:Sexpansion}
\end{equation}
where $\tilde{v}(z)=\tm_q z+\sigma z^3$ as in Eq.~(\ref{eq:X0}).
The terms in Eq.~(\ref{eq:Sexpansion}) with $z$-derivatives lead to boundary
terms in the variation of the action.
Expanding to quadratic order in $\tilde{S}$ and $\tilde{\pi}$, 
we find for the boundary variation of ${\cal S}$, \begin{multline}
\delta{\cal S}=\int_{\epsilon,z_m}d^4x\,{\rm Tr}\left\{\left[-\frac{2}{z^3}\partial_z \tilde{S}-
\frac{1}{z^3}\left(\tm_q+3\sigma z^2\right)\right]\delta\tilde{S} \right.
-\left.
\left[\frac{2(\tm_q z+\sigma z^3)^2}{z^3}\partial_z\tilde{\pi}\right]\delta\tilde{\pi}
\right\} . \end{multline}
To leading order in $\tm_q$ and in the fields,
using \begin{equation}
\delta\tilde{S}=-\tm_q z\left(\tilde{\pi}\,\delta \tilde{\pi}+
\delta \tilde{\pi}\,\tilde{\pi}\right)+\dots 
\label{eq:deltaStilde}\end{equation}
from Eq.~(\ref{eq:nonlinbc}), we find that the boundary variation
$\delta{\cal S}$ indeed vanishes.  
The cancellation of boundary variations in this nonlinear
fashion is novel in the context of extra-dimensional models, 
though it is reminiscent of the mixed boundary conditions of
certain Higgsless models \cite{Csaki:2003dt}
 in which the contributions to the boundary
variation of the action from different fields cancel one another.

In Kaluza-Klein theories an effective description of the lightest modes is often
derived by simply neglecting the heavier modes and integrating the action
over the extra dimension.  Indeed, that is how we derive the pion effective
action in this paper.  However, consistency of this approach relies on 
orthogonality and completeness relations dependent on the Sturm-Liouville
structure of the equations of motion and boundary conditions.  It is a mathematical
question which classes of systems of differential equations with nonlinear boundary
conditions satisfy the completeness and orthogonality theorems of Sturm-Liouville
systems.  In holographic QCD we have seen that 
there is a nonlinear field redefinition after which the boundary conditions are of
the linear Sturm-Liouville form.  This justifies the effective description 
obtained by keeping only the lightest modes.

\section{Couplings to Vector and Axial-Vector Fields}\label{sec:g5}
Having derived the chiral Lagrangian in the \(g_{5} \rightarrow 0\) limit,
we turn to the case of nonzero 5D gauge couplings with dynamical gauge bosons
representing the vector and axial vector mesons.
Including the gauge fields, the action takes the form of Eq.~(\ref{eq:5Daction}).
The field \(X\) transforms as a
bifundamental under the gauge group
\(SU(2)_{\mathrm{L}} \times SU(2)_{\mathrm{R}}\) and we will use the gauge
fixing condition \(L_{z}^{a} = R_{z}^{a} = 0\). We will also be working mainly
with the linear combinations \(A_{\mu}^{a} = (L_{\mu}^{a} -
R_{\mu}^{a})/2\), the axial vector field, and \(V_{\mu}^{a} = (L_{\mu}^{a} +
R_{\mu}^{a})/2\), the vector field.  The normalization of these combinations by a factor
of 2 (rather than $\sqrt{2}$) is so that their kinetic terms are canonically normalized
given the unconventional normalization of the gauge fields in Eq.~(\ref{eq:5Daction}).

We parameterize the fluctuations of the field \(X\) as in Eq.~(\ref{eq:Xnew}).
To leading order the scalars and pseudoscalars are decoupled, so we focus only on the 
pseudoscalars and set $S=0$ for the present discussion. 
As in the previous section, the boundary condition 
on the pion is modified as in Eq.~\eqref{eq:psibdy}.

We will determine the pion decay constant as in
Refs.~\cite{Erlich:2005qh,Da Rold:2005zs} by the residue of the
axial current two-point correlator at $q^2=0$.  The AdS/CFT calculation
of the correlator is performed by taking two functional derivatives, with respect
to the source of the axial current operator, of the action evaluated on the
classical solution to the linearized equation of motion for the transverse
part of \(A_{\mu}^{a}\). The resulting
correlator is in terms of the bulk-to-boundary propagator for \(A_{\mu}^{a}\),
which is a particular solution to the transverse-projected
linearized equation of motion.  This equation of motion for
\(A_{\mu}^{a}(q,z)_{\perp}\) is
\begin{equation}
\label{aperp}
\left[\partial_{z} \left(\frac{1}{z} \partial_{z} A_{\mu}^{a}\right)
  + \frac{q^{2}}{z} A_{\mu}^{a}
  - \frac{v^{2} g_{5}^{2}}{z^{3}} A_{\mu}^{a}\right]_{\perp} = 0.
\end{equation}
If we have a solution to Eq.~\eqref{aperp} of the form
\(A_{\mu}^{a}(q, z) = A(q, z) A_{0 \mu}^{a}(q)\), with boundary conditions
$\partial_z A(q,z)|_{z_m}=0$ and \(A(q, \epsilon) = 1\), then \(A(q, z)\) is
identified as the bulk-to-boundary propagator and \(A_{0\mu}^{a}(q)\) is the
source for the axial current.
The AdS/CFT prediction for the pion decay constant is then,
\begin{equation}
\label{eq:fpiAdSCFT}
  f_{\pi}^{2} = -\frac{1}{g_{5}^{2}} \left. \frac{\partial_{z} A(0,z)}{z}
    \right|_{z=\epsilon}.
\end{equation}

In order to study the pions we note that the pion fluctuations identified in the
field \(X\) mix with the longitudinal part of the axial vector field
\(A_{\mu}^{a} \rightarrow \partial_{\mu} \phi^{a}\), which has the same quantum
numbers. The pions will be identified as the lowest mode of this coupled
system. Since we have in mind a Kaluza-Klein decomposition of the fields and
since, for the purposes of deriving the low energy theory, we are only
interested in the lowest mode,
we will make the substitutions \(\pi^{a}(x, z) \rightarrow \pi^{a}(x) \psi(z)\)
and \(\phi^{a}(x, z) \rightarrow \pi^{a}(x) \phi(z)\).
The linearized equations of motion for \(\psi(z)\) and \(\phi(z)\) are
\begin{equation}
\label{psiphieom}
\begin{aligned}
  v \phi - \sigma z^{3} \psi =
    \frac{z^{3}}{v g_{5}^{2}} \partial_{z} \left(\frac{1}{z} \partial_{z}
    \phi \right),
\\
  m_{\pi}^{2} (v \phi - \sigma z^{3} \psi)
    = \partial_{z} \left(\sigma z^{3} \partial_{z} \psi \right),
\end{aligned}
\end{equation}
where the fields satisfy the boundary conditions \(\partial_{z}\phi(z)|_{z_{m}}
= \phi(\epsilon) = \psi(\epsilon) = 0\), and Eq.~\eqref{eq:psibdy}.

\subsection{Approximate Analytic Results}
We can obtain
an approximate solution to the equations of motion, Eqs.~\eqref{psiphieom}, in
the chiral limit, in
a similar fashion to Ref.~\cite{Erlich:2005qh}.  We find that the approximate
solutions near the \(\epsilon\) boundary are \(\phi(z) = 1 - A(0,z)\) and
\(\psi(z) = 0\), while those away from
the \(\epsilon\) boundary are \(\phi(z) = 1 - A(0, z)\) and
\(\psi(z) \approx 1\). Plugging the first equation of Eq.~\eqref{psiphieom}
into the second, approximating $v(z)\approx \sigma z^3$, 
and integrating once we arrive at
\begin{equation}
  \frac{m_{\pi}^{2}}{\sigma g_{5}^{2}}\left(\frac{1}{z}\partial_{z}\phi
    - \left. \frac{1}{z}\partial_{z} \phi \right|_{\epsilon}\right)
    = \sigma z^{3} \partial_{z} \psi.
\end{equation}
Now if we evaluate this expression on the IR boundary, using our approximate
solution for \(\phi(z)\) and recalling the boundary conditions, we find
\begin{equation}
  m_{\pi}^{2} \left(-\frac{1}{g_{5}^{2}} \left. \frac{\partial_{z} A(0,z)}{z}
      \right|_{\epsilon}\right) = 2 m_{q} \sigma \psi(z_{m}).
\end{equation}
By utilizing Eq. \eqref{eq:fpiAdSCFT} and the fact that
\(\psi(z_{m}) \approx 1\), we have once again derived the GOR relation.

The derivation of the chiral Lagrangian mass term is similar to that of
previous sections.  The only contributions come from the term proportional
to \(\partial_{z}U \partial_{z}U^{\dagger}\) and the boundary term proportional
to \(U + U^{\dagger}\).  In particular we have, upon integration by parts,
\begin{equation}
\begin{split}
{\cal S} &\supset \int d^{4}x \left[\int dz \,\partial_{z}
    \left(\sigma z^{3} \partial_{z} \psi\right) \psi
  + 2 m_{q} \sigma \psi(z_{m})^{2} \right] \frac{1}{2} \pi^{a}(x) \pi^{a}(x)
  \\ &\quad
  + \int d^{4}x \, {\rm Tr} \left\{\frac{m_{q} \sigma}{2} (U
  + U^{\dagger})\big|_{z_{m}}\right\}.
\end{split}
\end{equation}
If we make use of the equations of motion, Eq.~\eqref{psiphieom}, and substitute
our approximate solutions for \(\phi\) and \(\psi\), we find that the two
terms in the square brackets cancel, to first order in \(m_{q}\). Thus,
to this order in \(m_{q}\), we are left with the mass term of the
chiral Lagrangian:
\begin{equation}
{\cal S}_{\mathrm{4D}} \supset \int d^{4}x \, {\rm Tr}
    \left\{\frac{m_{\pi}^{2} f_{\pi}^{2}}{4} (U + U^{\dagger})\right\}.
\end{equation}

\subsection{Numerical Results}
\begin{figure} [ht]
  \centering
  \mbox{\subfigure[]{\includegraphics[scale=0.3]{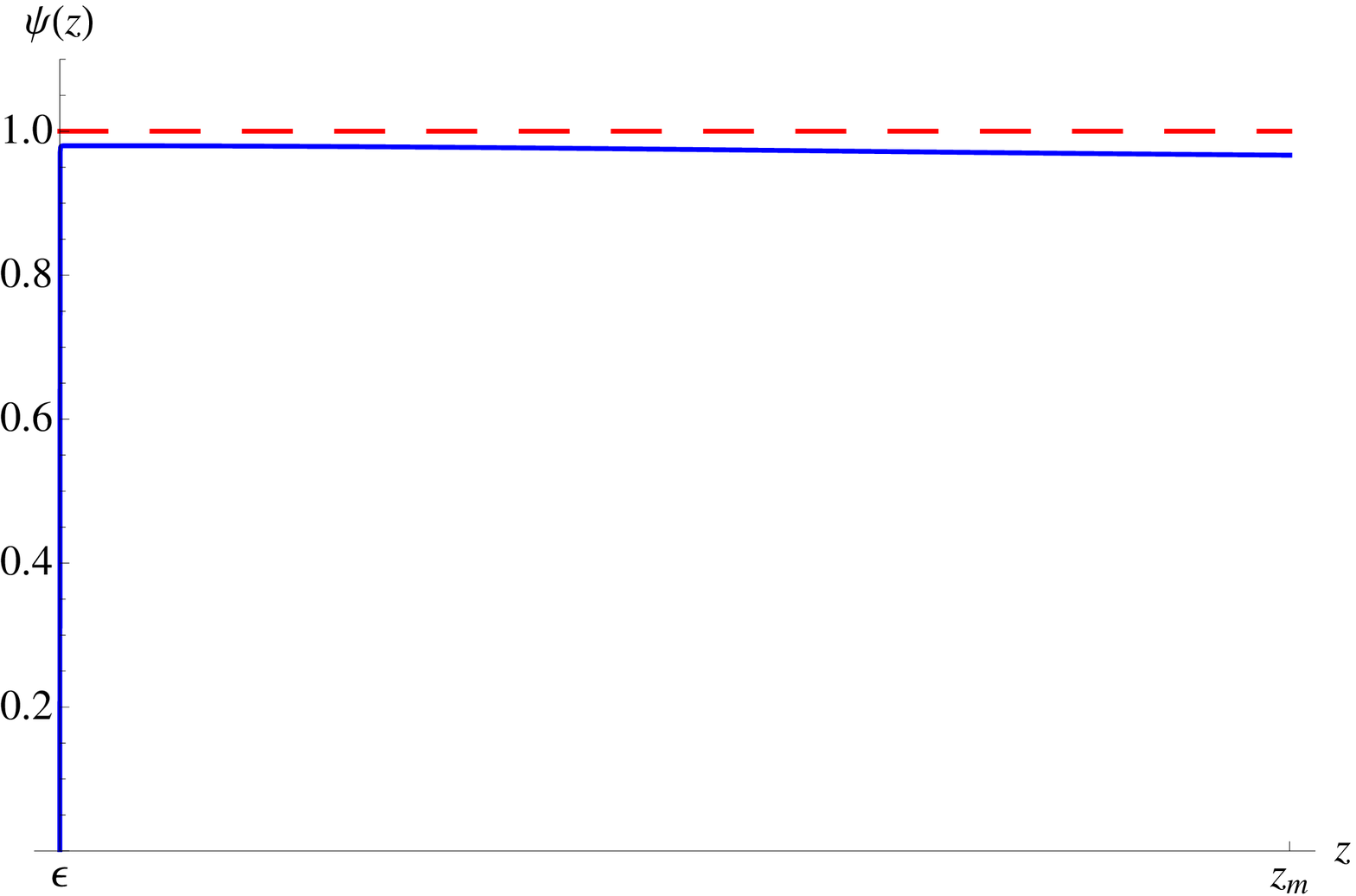}} \quad
    \subfigure[]{\includegraphics[scale=0.3]{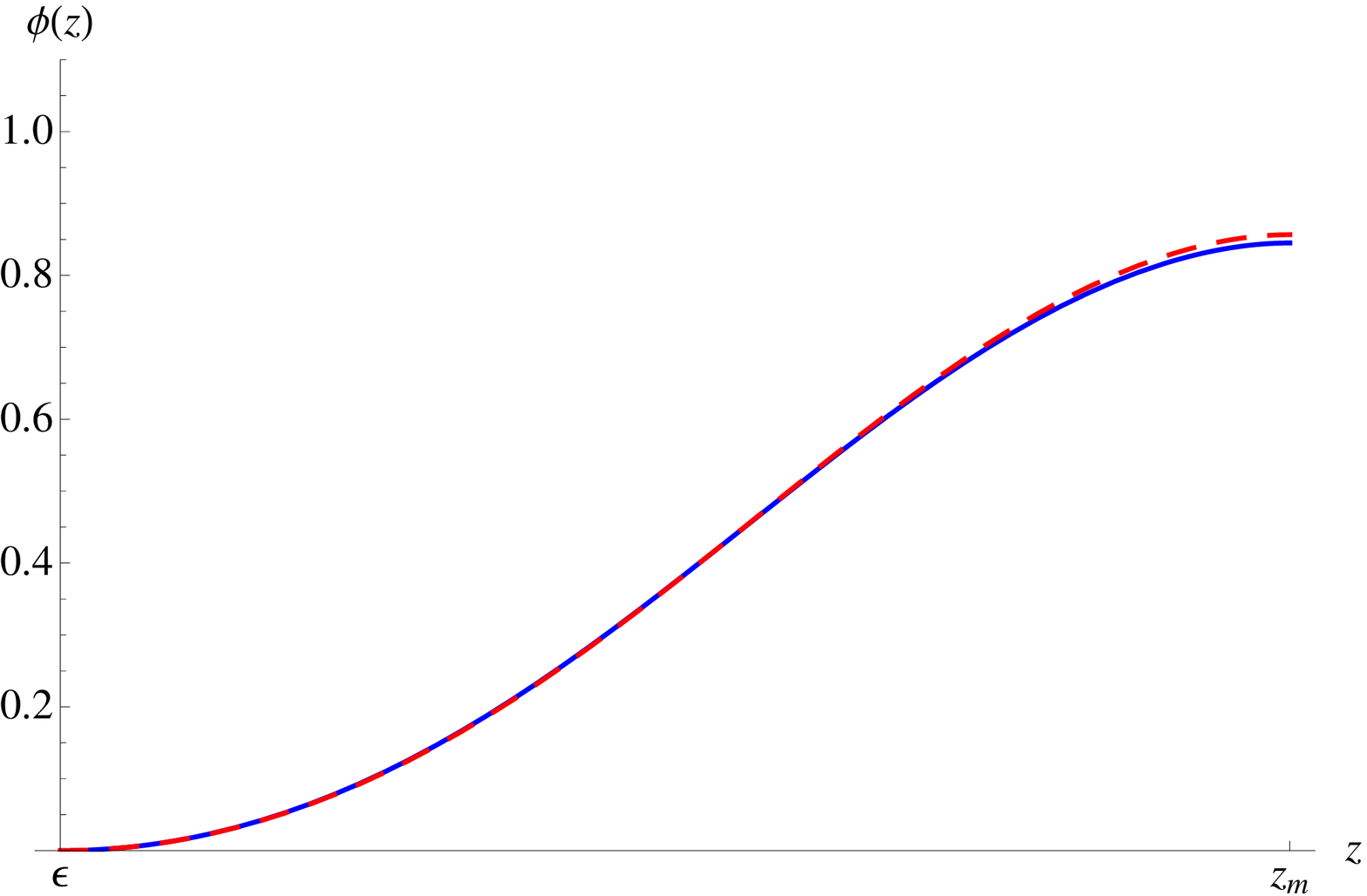}}
  }
  \caption{Figures (a) and (b) display the numerical solutions to
    Eq.~\eqref{psiphieom} for \(\psi(z)\) and \(\phi(z)\), respectively.
    In both figures there are two curves plotted: the blue curves are the
    numerical solutions and the red, dashed curves are the approximate
    solutions (\(\psi(z) = 1\) in figure (a) and \(\phi = [1 - A(0,z)]\) in
    figure (b)).  These plots were made with \(m_{q} = 2.36\ {\rm MeV}\) and
    \(\sigma = (333\ {\rm MeV})^{3}\).}
  \label{fig:numprofiles}
\end{figure}

We will now present a numerical analysis of the equations of motion,
Eqs.~\eqref{psiphieom}.  We choose a
value \(\epsilon = 10^{-7}\ {\rm MeV}\) for the UV cutoff, and determine the
location of the IR boundary to be \(z_{m} = 1/(323\ {\rm MeV})\), by setting
the rho mass to \(776\ {\rm MeV}\) \cite{Erlich:2005qh}. We take
\(g_{5} = 2 \pi\) as in
Ref.~\cite{Erlich:2005qh}, noting that the derivation of this assignment is
unaffected by our new choice for the form of the field \(X(x,z)\), with
different background.   With the
values \(m_{q} = 2.36\ {\rm MeV}\) for the quark mass and \(\sigma = (333\
{\rm MeV})^{3}\) for the condensate, we have \(m_{\pi} = 140\ {\rm MeV}\)
for the pion mass and \(f_{\pi} = 92.0\ {\rm MeV}\) for the pion decay constant.
The solutions for \(\psi(z)\) and \(\phi(z)\) are plotted in
Fig.~\ref{fig:numprofiles}, along with their approximate solutions, namely
\(\psi(z) = 1\) and \(\phi(z) = 1 - A(0, z)\).
The functions \(\psi(z)\) and \(\phi(z)\) are normalized to obtain a canonically
normalized kinetic term in the low energy theory.  The plotted numerical
solutions illustrate the extent to which the approximations of the previous
section are valid.

\begin{figure} [h!]
  \centering
  \includegraphics[scale=0.3]{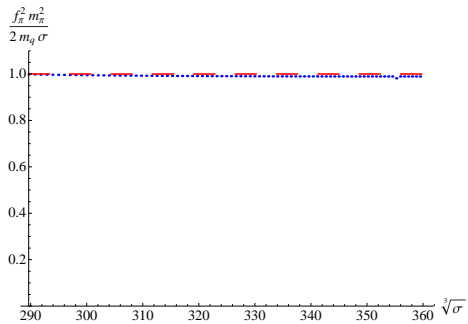}
  \caption{The red long-dashed line is a horizontal line at 1.  The blue dots
    are the values of the function, \((m_{\pi}^{2} f_{\pi}^{2}) / (2 m_{q}
    \sigma)\).  For this plot, the quark mass is adjusted to
    fix \(m_{\pi} = 140\ {\rm MeV}\) at each point.}
  \label{fig:gor}
\end{figure}

We would also like to understand numerically how robust the GOR relation is in
this model, with respect to varying some of the parameters.  
In particular, fixing \(m_{\pi} = 140\ {\rm MeV}\) (by
adjusting \(m_{q}\) for fixed values of $\sigma$) 
we would like to see what happens for different
values of \(\sigma\). Varying the condensate by sampling a discrete number
of points from \((290\ {\rm MeV})^{3}\) to
\((360\ {\rm MeV})^{3}\), the pion decay constant takes values ranging from
\(79\ {\rm MeV}\) to \(102\ {\rm MeV}\) and the quark mass goes from
\(2.52\ {\rm MeV}\) down to \(2.21\ {\rm MeV}\).  In Fig.~\ref{fig:gor},
we plot the ratio \((m_{\pi}^{2} f_{\pi}^{2}) / (2 m_{q} \sigma)\) for the
above specified range of values for the condensate.  If the GOR relation holds,
the ratio should be approximately 1 and we can see that the plot shows good
agreement over the entire region.

\section{Consequences for Holographic QCD}
We check that the change of background does not adversely affect some standard
predictions of the hard-wall model; that is, we will compare our predictions
with those of ``Model A" in Ref.~\cite{Erlich:2005qh}.
We might expect the modifications to be marginal because only the pion physics
should be sensitive to changes in \(m_{q}\), but we have also changed a sign.
Aside from the substitution \(\tm_{q} \rightarrow -2 m_{q}\), the derivations
of the equations of motion for the gauge fields and of the vector current
correlators are unaffected by the new form of the \(X\)-field.
Since the expressions for the related observables we compute are likewise
unchanged, we refer the reader to Ref.~\cite{Erlich:2005qh}, for more detail.
In the following, we summarize the methods for obtaining the results. To
calculate the mass of the \(a_{1}\), we solve Eq.~\eqref{aperp} for the
normalizable mode \(\psi_{a_{1}}(z)\) with boundary conditions
\(\psi_{a_{1}}(\epsilon) = \partial_{z}\psi_{a_{1}}(z_{m}) = 0\). The wavefunction
$\psi_{a_1}$ is normalized such that $A_\mu(x)$ has a canonical kinetic term,
where $A_\mu(x,z)=\psi_{a_1}(z)A_\mu(x)$ .
As derivable
from the two-point vector current correlator, we use the following expression
to calculate the decay constants in terms of the profile in the extra dimension:
\(F^{1/2} = \sqrt{\psi''(0)/g_{5}}\).
And finally, by looking at the terms cubic in the fields, coupling \(V \pi \pi\),
we make a prediction for \(g_{\rho \pi \pi}\).
All relevant cubic terms are the following:
\begin{equation}
{\cal S} \supset \epsilon^{abc} \int d^{5}x\ V_{\mu}^{b}
    \left[\frac{1}{z g_{5}^{2}} \partial_{z} \partial^{\mu}\phi^{a}
    \partial_{z}\phi^{c}
   + \frac{1}{z^{3}} \partial^{\mu}\left(v \phi^{a}
   - \sigma z^{3} \pi^{a} \right)
     \left(v \phi^{c} - \sigma z^{3} \pi^{c} \right)
    \right],
\label{vpipi}
\end{equation}
where we have used the equations of motion for the vector field to obtain this
expression. In order to calculate the on-shell \(g_{\rho \pi \pi}\)
from the effective 4D theory, we integrate
out the extra dimension and identify \(g_{\rho \pi \pi}\) as the coefficient of
\(\epsilon^{abc} \partial^{\mu} \pi^{a}(x) \rho_{\mu}^{b}(x) \pi^{c}(x)\),
where \(\rho_{\mu}^{b}\) is the lowest mode in the vector field KK decomposition.
Extracting this coefficient from Eq.~\eqref{vpipi} we find
\begin{equation}
g_{\rho \pi \pi} = \frac{g_{5}}{f_{\pi}^{2}} \int dz\ \psi_{\rho}(z)
    \left[\frac{1}{z g_{5}^{2}} (\partial_{z} \phi)^{2}
   + \frac{1}{z^{3}} \left(v \phi - \sigma z^{3} \psi\right)^{2}
    \right],
\end{equation}
where the \(z\)-integral of the expression in brackets is normalized to
\(f_{\pi}^{2}\).  The results are presented in Table \ref{table:results} and
we find that the new predictions have not changed significantly
compared to ``Model A" -- they are still on the \(10\%\) level, with the
exception of \(g_{\rho \pi \pi}\).
\begin{table}[ht]
\begin{ruledtabular}
\begin{tabular}{ccrr}
  Observable & Measured (MeV) & Model (MeV) & Model A (MeV)\cite{Erlich:2005qh}\\
\hline
  \(m_{\pi}\) 		& 140 	& 140		& 140\\
  \(m_{\rho}\) 		& 776 	& 776		& 776\\
  \(m_{a_{1}}\) 	& 1230 		& 1370		& 1360\\
  \(f_{\pi}\) 		& 92.4 		& 92.0		& 92.4\\
  \(F_{\rho}^{1/2}\) 	& 345 		& 329		& 329\\
  \(F_{a_{1}}^{1/2}\) 	& 433 		& 493		& 486\\
  \(g_{\rho \pi \pi}\) 	& 6.03 		& 4.44		& 4.48
\end{tabular}
\end{ruledtabular}
\caption{\label{table:results}The predictions of the model as compared with
``Model A" and experimental values \cite{Erlich:2005qh}.  The results are
based on fitting to \(m_{\pi}\), \(f_{\pi}\), and \(m_{\rho}\), leading to
the parameter choice \(m_{q} = 2.36\ {\rm MeV}\),
\(\sigma = (333\ {\rm MeV})^{3}\), and \(z_{m} = 1/(323\ {\rm MeV})\).}
\end{table}

\section{Conclusions}
Holographic QCD models are surprisingly successful in their predictions of
low-energy QCD observables.  However, it was discovered in earlier work that the pion 
condensation transition in one version of the hard-wall model 
has qualitatively different features than
the predictions of chiral perturbation theory and other approaches.  
We have shown that this
disagreement is due to a boundary effect related to the explicit breaking of the
gauged chiral symmetry by the non-normalizable background of a 5D scalar field.  
To restore agreement with the chiral Lagrangian we modified the boundary dynamics,
either by introducing nonlinear boundary conditions on the fields, or by
performing a nonlinear
field redefinition which induced an infrared boundary term in the action.
The field redefinition allowed us to relate the system with nonlinear boundary conditions
to a standard Sturm-Liouville system which manifestly maintains the proper
symmetry structure, justifying the subsequent Kaluza-Klein decomposition
of the fields.
The chirally improved hard-wall model makes predictions for low-energy QCD observables that
agree with the original model to within 1-2\%.

It would be useful to further explore the
consequences of the modified boundary physics with regard to pion observables.  
It would also be interesting to find additional applications of nonlinear
boundary conditions to extra-dimensional model building, for example to Higgsless
models and holographic technicolor models. 
Finally, the necessity for nonlinear boundary dynamics in the hard-wall model
provides motivation for further study of the 
mathematical problem of differential equations with nonlinear boundary conditions.
In particular, it would be useful to classify those systems of equations and boundary
conditions that can be related to a Sturm-Liouville system by a change of variables.

\begin{acknowledgments}
We are happy to thank Chris Carone, Tom Cohen, 
Will Detmold, Stefan Meinel and Reinard Primulando for useful discussions.
This work was supported by the NSF under Grants PHY-0757481 and PHY-1068008.
\end{acknowledgments}

\end{document}